\documentclass[a4paper,11pt]{article}

\usepackage{amssymb,amstext,amsmath,amsthm}
\usepackage{graphicx}
\usepackage{latexsym}
\usepackage{psfrag}
\usepackage{amsfonts}
\usepackage{verbatim}
\usepackage{vmargin}
\usepackage{helvet}

\usepackage{amsmath,amsxtra,amsthm,amssymb,xr}
\usepackage[all]{xy}
\usepackage[]{mathrsfs}

\usepackage{color}

\newcommand{\bea}{\begin{eqnarray}}
\newcommand{\eea}{\end{eqnarray}}
\newcommand{\nn}{\nonumber}
\newcommand{\be}{\begin{equation}}
\newcommand{\ee}{\end{equation}}

\newcommand{\nb}{{\bf n}}

\newcommand{\R}{\mathbb{R}}
\newcommand{\C}{\mathbb{C}}

\newcommand{\lalg}[1]{\mathfrak{#1}}  
\newcommand{\SU}{\mathrm{SU}}
\newcommand{\SO}{\mathrm{SO}}

\newcommand{\U}{\mathrm{U}}
\newcommand{\Spin}{\mathrm{Spin}}
\newcommand{\Hom}{\mathrm{Hom}}
\newcommand{\Reel}{\mathrm{Re}}
\newcommand{\su}{\lalg{su}}

\newcommand{\so}{\lalg{so}}

\def\la{\langle}
\def\ra{\rangle}

\setmarginsrb{2.3cm}{.9cm}{2.3cm}{1.8cm}{12pt}{11mm}{0pt}{13mm}

\title{A Summary of the asymptotic analysis for the EPRL amplitude }
\author{John W. Barrett\footnote{john.barrett@nottingham.ac.uk}
, Richard J. Dowdall\footnote{richard.dowdall@maths.nottingham.ac.uk}
, Winston J. Fairbairn\footnote{winston.fairbairn@nottingham.ac.uk}
, \\ Henrique Gomes\footnote{henrique.gomes@maths.nottingham.ac.uk}
, Frank Hellmann\footnote{frank.hellmann@maths.nottingham.ac.uk}
}

\begin{document}

\maketitle

\begin{abstract}
We review the basic steps in building the asymptotic analysis of the Euclidean sector of new spin foam models
using coherent states, for Immirzi parameter less than one. We  focus on conceptual issues and by so doing omit
 peripheral proofs and the original discussion on spin structures.
\end{abstract}
\section{Introduction}
The present work consists in the report of a talk given by one of us (HG) in the Planck Scale 2009 Conference,
which took place in Wroclaw, and is based entirely on \cite{Barrett:2009gg}.

 A spin foam model \cite{perez-2003-20} is a  procedure to compute an
amplitude from a triangulated manifold $\mathcal{T}$ with $n$-simplices $\Delta_n$ coloured by representation
theory data. In four-dimensions, such an amplitude is typically of the form \be\label{spinfoam}
\mathcal{Z}(\mathcal{T}) = \sum_{\iota, \rho} \prod_{\Delta_2} f_2 (\rho) \prod_{\Delta_3} f_3(\rho,\iota)
\prod_{\Delta_4} f_4(\rho,\iota)~, \ee where $f_n$ are weights assigned to the $n$-simplices of the triangulated
manifold, and $\rho$ and $\iota$ respectively denote the assignments of unitary, irreducible representations to
the $2$-simplices, and intertwining operators to the $3$-simplices of $\mathcal{T}$. The model is specified by the
choice of representation assignments, the vector space of intertwining operators $\iota$, and weights $f_n$.

A key step in understanding the semiclassical  regime of a spin foam model
 in dimension $d$ is the analysis of the asymptotic behaviour of
the $d$-simplex amplitude that defines the model. For instance, what really established the Ponzano-Regge spin
foam model as a model for 3D quantum gravity \cite{ponzanoregge} was the discovery that it had some very tangible
geometric interpretation, {\it in the asymptotic limit}. The
discovery by Ponzano and Regge that it contains the geometry of the tetrahedron through the Regge action was the
crucial step for the corresponding spin foam model.

 Similar asymptotic analysis of the 4-dimensional models
\cite{barrett-1998-39} was initially performed by Barrett and Williams
\cite{barrett-1999-3}, and formed the basis of investigations of the
graviton propagator structure of these models \cite{Alesci:2008gv}. This latter
analysis showed a definite incompatibility between the 10j symbol and a boundary structure given by loop quantum
gravity-like geometry. Consequently a host of new 4-dimensional models were developed. We here will discuss only
a refined version of the original EPR model written with
  Livine (EPRL) \cite{Engle:2007wy}, in the case $\gamma<1$.

\section{Briefly introducing the EPRL model}
As is well known, even though GR has local degrees of freedom, it can  be put into a `BF shape' by the use of the
action:
$$S_{GR}=\int_M\mbox{tr}\left( *(
 e\wedge e) \wedge F(A)\right)$$
for the Lie algebra valued two forms $F(A)=dA+A\wedge A$ and $B = * (e\wedge e)\in\Lambda^2(M,\so(4))$, where $e$
denotes the (co)frame field $e\in\Lambda^1(M,\R^4)$ (we use the identification $\Lambda^2( \R^4)\simeq\so(4)$ )
and $*$ is the $\so(4)$ Hodge. By restricting the sum over representations and intertwiners in the BF partition
function to respect this constraint on the $B$ field, Barrett and Crane derived their 4-dimensional spin foam
model \cite{barrett-1998-39}.

A host of new 4-dimensional models have been recently developed, based on the classically equivalent Holst action:
$$S_{GR}=\int_M \mbox{tr} \left((* (e\wedge e) + \frac{1}{\gamma} e \wedge e) \wedge
 F(A)\right)$$
where $\gamma$ is the so called Immirzi parameter, and the restrictions on  the representations and intertwiners
are of different form from the BC model. Namely, the EPRL allowed representations are constructed from the
 Clebsch-Gordan decomposition:
  $$V_{j^-} \otimes V_{j^+} \simeq
\bigoplus^{j^++j^-}_{k=|j^--j^+|}V_{k} $$ One then takes the projection onto the highest weight: $k = j^+ +
j^-$, and forms the Clebsch-Gordan intertwining map $C^{j^- j^+}_{k} \colon V_k\rightarrow V_{j^-} \otimes V_{j^+}
$
 injecting  into the highest (diagonal $\SU(2)$ subgroup) factor. The labels $j^{\pm}$ and $k$ are related via the
Immirzi parameter for $\gamma<1$ by \be j^{\pm} = \frac{1}{2} (1 \pm \gamma) \, k. \ee This tells us a specific
way to go from an $\SU(2)$ irrep to a tensor product of two $\SU(2)$ irreps, i.e. to a $\Spin(4)$ irrep.

Moving on to the intertwiners, an $\SU(2)$ intertwiner $\hat{\iota}$ is an element of
$\Hom_{\SU(2)}(\C,\bigotimes_{i=1}^4 V_{k_i})$. From the above construction of the injection of irreps of $\SU(2)$
into those of  $\Spin(4)$,  and given an $\SU(2)$ intertwiner $\hat{\iota}$, a $\Spin(4)$ intertwiner $\iota$ is
constructed as follows:
\begin{figure}
\begin{center}
\psfrag{a}{$ \int_{\Spin(4)} \, dG $}
\psfrag{b}{$\hat{\iota}^{k_1 k_2 k_3 k_4}$}
\psfrag{j1+}{$ j^+_{1}$}
\psfrag{j1-}{$ j^-_{1}$}
\psfrag{j2+}{$ j^+_{2}$}
\psfrag{j2-}{$ j^-_{2}$}
\psfrag{j3+}{$ j^+_{3}$}
\psfrag{j3-}{$ j^-_{3}$}
\psfrag{j4+}{$ j^+_{4}$}
\psfrag{j4-}{$ j^-_{4}$}
\psfrag{k1}{$ k_1$}
\psfrag{k2}{$ k_2$}
\psfrag{k3}{$ k_3$}
\psfrag{k4}{$ k_4$}
\psfrag{g+}{$ X^+$}
\psfrag{g-}{$ X^-$}
\psfrag{c1}{$ C^{j_1^+ j_1^-}_{k_1}$ }
\psfrag{c2}{$ C^{j_2^+ j_2^-}_{k_2}$ }
\psfrag{c3}{$ C^{j_3^+ j_3^-}_{k_3} $}
\psfrag{c4}{$ C^{j_4^+ j_4^-}_{k_4} $}
\includegraphics[scale=0.4,trim=0mm -10mm -40mm -10mm]{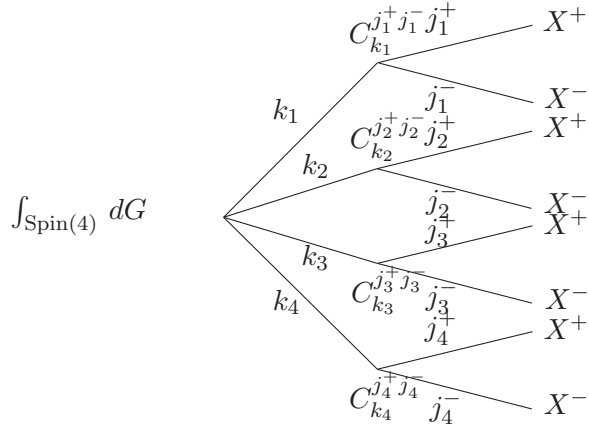}
\caption{The $\Spin(4)$ intertwiner $\iota$.}\label{figone}
\end{center}
\end{figure}
\be\label{intertwiner} \iota := \int_{\Spin(4)} \, dG \,\, ({j_i^-} \otimes {j_i^+})(G) \circ \bigotimes_{i=1}^4
C^{j_i^- j_i^+}_{k_i} \circ \hat{\iota} \,\, , \ee where the notation $G=(X^-, X^+)\in\Spin(4)=\SU(2)\times\SU(2)$
is used (see figure \ref{figone}).  The group integration ensures that the resulting object is
$\Spin(4)$-invariant, i.e., is an element of $\Hom_{\Spin(4)}(\C,\bigotimes_{i=1}^4 V_{(j_i^-,j_i^+)})$.

Labelling the tetrahedra by $a=1,...,5$, the ten triangles $\Delta_2$ of the 4-simplex $\Delta_4$ are then indexed
by the pair $ab$ of tetrahedra which intersect on the triangle. There are two $\SU(2)$ group elements
$(X_a^-,X_a^+)$ and one $\SU(2)$ intertwiner $\hat{\iota}_a$ for each tetrahedron. The above $\Spin(4)$
intertwiners are glued together in the standard fashion (the usual pentagon combinatorics) to construct an
amplitude (a complex number) for this data. Note that now the input data for this  4-simplex $\Spin(4)$  amplitude
is  a spin $k\in\{0,\frac12,1,\ldots\}$ for each triangle of the 4-simplex and an $\SU(2)$ intertwiner
$\hat{\iota}$ for each tetrahedron.

 This concludes the basic construction of the
model. We  opt for not writing the amplitude in the present form, preferring to first write it into a form
appropriate for asymptotic analysis. Nonetheless, one can see that the asymptotic problem could not yet be well posed,
because the scaling of the $\SU(2)$ intertwiners is not  defined in the present form. Solving this problem
naturally leads to a reformulation of the integral formula to an exponential form which is particularly well
suited to asymptotics.
\section{Coherent: states and tetrahedra}

\paragraph*{States:} The  fundamental new tool that permitted the asymptotic analysis of the new models was the
introduction of {\it coherent states}. Heuristically, these are states of some irrep of $\SU(2)$ that are  most
geometrical, or semi-classical in the sense that they minimize the uncertainty in total angular momentum
\cite{livine-2007-76}. The coherent states have  maximal spin projection along the $\mathbf n$ axis, i.e. they are
highest weight eigenvectors of the normalized Lie algebra elements. Explicitly, for $L^j = \frac{i}{2}\sigma^j$
the Lie algebra generators and $\nb \in S^2$, a coherent state $|k,{\bf n}\ra \in V_k$ in direction $\mathbf n$ is
a unit vector satisfying \be\label{coherent states}(\mathbf L.\mathbf n) \, |k,{\bf n}\ra = i k \, |k,{\bf
n}\ra\ee where the dot `.' denoting the 3d (Euclidean) scalar product. At each point, there is a $\U(1)$ family of
coherent states that satisfy \eqref{coherent states}, and we have denoted a fixed initial arbitrary choice
 as\footnote{For fixed $k$ this is equivalent to a section of the Hopf bundle, $
s:S^2\simeq\SU(2)/\U(1)\rightarrow\SU(2)$. Locally we can denote any other choice by $e^{i\theta({\bf n})}|k,{\bf
n}\ra$.} $|k,{\bf n}\ra$.

Coherent states have the following properties which will be useful to us:
  \begin{enumerate}\label{Properties}\item  $ g |k,{\bf n}\ra=e^{ik\phi}|k,\hat g{\bf n}\ra$ where
  $g\in\SU(2)$, with $\SO(3)$ projection $\hat g$,  and $\phi$ is an arbitrary phase. This means that the
  action of $\SU(2)$ takes a coherent state for a vector ${\bf n}$ into a coherent state for a vector $\hat
  g{\bf n}$.
\item  $|k,{\bf n}\ra = |\frac{1}{2},{\bf n}\ra^{\otimes 2k}=:|{\bf n}\ra^{\otimes 2k}$ so coherent states exponentiate into the fundamental
representation, which in diagrammatical calculus means we replace a strand labelled $k$ by $2k$ {\it
identical} fundamental strands.
\end{enumerate}

\paragraph*{Tetrahedra:} To construct a coherent intertwiner associated to a tetrahedron, the idea is to
associate a coherent state to each one of its triangles and then integrate over $\SU(2)$.
  The geometrical picture is that the coherent
intertwiner corresponding to tetrahedron $a$ of the triangulation will be given by a `coherent tetrahedron'
 labeled $\tau_a$.
  Here $\tau_a$ has a coherent state $|k_{ab}, \mathbf n_{ab} \ra$ for each face, carrying  the
interpretation of the normals of length $k$ and direction $\mathbf n_{ab}$ (and an implicit choice of
 phase factor). Thus apart from the phase factor, we can in effect regard $\tau_a$ as a tetrahedron in $\R^3$
 with the standard metric, with
 $\mathbf n_{ab}$ and $k_{ab}$ being the normal and area associated to $\tau_{ab}\subset\tau_a$; the triangle of $\tau_a$
 (combinatorically) adjacent to tetrahedron $b$.

Of course, we want to describe  tetrahedra with three-dimensional rotational symmetry, so the coherent
intertwiners are  constructed by integrating over all spatial directions the tensor product of four coherent
states \be\label{intertwiner2} \hat{\iota}(\mathbf n_1,\mathbf n_2,\mathbf n_3,\mathbf n_4) = \int_{\SU(2)} dh
\,\bigotimes_{i=1}^4 h | k_i, \mathbf n_i \ra . \ee These intertwiners were introduced by Livine and Speziale
\cite{livine-2007-76}, who gave an asymptotic formula for their normalisation.

According to the `quantization commutes  with reduction' theorem of Guillemin and Sternberg \cite{GS}, the space
of intertwiners is spanned by the $\hat \iota$ determined by vectors satisfying the {\em closure constraint} $
k_1 \mathbf n_1+ k_2 \mathbf n_2+k_3\mathbf n_3+k_4\mathbf n_4=0$. Thus we take the coherent
intertwiners to always satisfy this condition and thus be given by some tetrahedron $\tau$.

Given the above formulation of the coherent intertwiner for $\SU(2)$,  we know from equation \eqref{intertwiner}
what the form of the $\Spin(4)$-intertwiner should be.\footnote{Note that the invariance of the Clebsch-Gordan
injections permits us to absorb the  $\SU(2)$ integration of the coherent tetrahedra into the $\Spin(4)$
integration.}

\section{Exponential form and stationary points}
\paragraph*{Writing the amplitude in exponential form}
The amplitude $f_4\in\C$  is defined by forming a closed spin network diagram from the 5
 $\Spin(4)$
intertwiners (vertices) $\iota_{a}$, which are tensored together and then the free ends are joined pairwise
according to the combinatorics. This is done using the standard `$\epsilon$ inner product' of irreducible
representations of $\SU(2)$, denoted $\epsilon_k\colon V_k\otimes V_k\to\C.$
 This inner product is represented in the spin network
diagram as a semicircular arc\footnote{This is defined by a choice of the two-dimensional antisymmetric tensor
$\epsilon$ for  $\SU(2)$ spin $1/2$, and extended to arbitrary spin by tensor products of $\epsilon$. This choice
of  inner product  makes the combinatorics and $-1$ signs tractable, but  also allows the natural assignment of
coherent states (and hence intertwiners) to the triangle normals.}. To toggle  between the usual Hermitian inner
product and the epsilon inner product,
 one uses the standard antilinear structure map for representations
of $\SU(2)$, $J\colon V_k\to
V_k.$ This is defined by
$$\epsilon_k(\alpha,\alpha')= \la J \, \alpha | \alpha' \ra,$$
the left-hand side being the epsilon-inner product and the right hand side the Hermitian inner product. It obeys
$Jg=gJ$ for all $g\in\SU(2)$, $J^2=(-1)^{2k}$ and $\langle
J\alpha|J\alpha'\rangle=\overline{\langle\alpha|\alpha'\rangle}$. Furthermore, since
$ J(i \mathbf n\cdot\mathbf L)= - (i \mathbf n\cdot \mathbf L) J,$
the map $J$ takes a coherent state for ${\bf n}$ to a coherent state for $-{\bf n}$, hence the notation
$|k_{ab},-{\bf n}_{ab}\ra$ means $J|k_{ab},{\bf n}_{ab}\ra$.

To combine the $\Spin(4)$ intertwiners $\iota_{a}$, one first of all regards each vertex as an $\SU(2)$ spin
network (as in figure \ref{figone}),  and uses one $\epsilon$ inner product to connect the $j^+$ edges and a
second $\epsilon$ inner product to connect the $j^-$ edges. Using the form \eqref{intertwiner2} for the
intertwiners, one splits the total amplitude into a $\Spin (4)^5$ integral, where the integrand is a product of
spin network evaluations, one for each edge of the 4-simplex (see also footnote 2). We call these evaluations {\it
propagators}, denoted by $ \mathcal{P}_{ab}$.

It is easy to see that the symmetrizers on the $j^+_{ab}$ and $j^-_{ab}$ edges can be absorbed into the
symmetrizer on the $k_{ab}$ edge because of the stacking property of symmetrizers. Furthermore, using the
exponentiating property of the coherent states, the remaining symmetrizer
  now acts redundantly on the coherent  states $ | k_{ab}, \mathbf n_{ab} \rangle$, and we can
further split the propagator into the $j^+$ and $j^-$ strands, i.e. to a product of terms in the fundamental
representation. We obtain the following expression for the propagator \be \mathcal{P}_{ab} = \la -\mathbf n_{ab}|
(X^-_a)^{-1}X^-_b |  \mathbf n_{ba} \ra^{2j_{ab}^-} \; \la -\mathbf n_{ab}| (X^+_a)^{-1}X^+_b | \mathbf n_{ba}
\ra^{2j_{ab}^+}. \ee  The four-simplex amplitude can thus be re-expressed as $ f_4 =  \int_{\Spin(4)^5} \prod_{a}
dG_a \; e^{S}$ with the action given by \be \label{action} S = \sum_{a < b} \, 2j_{ab}^- \, \ln \, \la -\mathbf
n_{ab}| (X^-_a)^{-1}X^-_b |   \mathbf n_{ba} \ra + 2j_{ab}^+ \, \ln \, \la -\mathbf n_{ab}| (X^+_a)^{-1}X^+_b |
\mathbf n_{ba} \ra. \ee

\paragraph*{Stationary Points}
We start by scaling all ten spins by a constant parameter $k_{ab} \rightarrow \lambda k_{ab}$. Our strategy is to
use extended stationary phase methods, that is, stationary phase generalized to (non purely imaginary) complex
functions, to find, in terms of the boundary data,  the $\Spin(4)$ elements $G_a=(X_a^-,X_a^+)$ that leave the
action stationary.

In the extended stationary phase, the key role is played by {\em  critical points}, i.e {stationary points} for
which $\Reel S=0$. If $S$ has no critical points then
  for large
parameter $\lambda$ the function $f$ decreases faster than any power of $\lambda^{-1}$. In other words, for all $N
\geq 1$: $ f(\lambda) = o(\lambda^{-N})$. Otherwise, for large $\lambda$ the asymptotic expansion of the integral
$f(\lambda) = \int_D d x \, a(x) \, e^{\lambda S(x)}$ yields for each critical point \cite{Hormander}\footnote{The
stationary points are assumed to be isolated and non-degenerate; $\det H \neq 0$} $$ \label{formula1}  a(x_0)
\left(\frac{2 \pi}{\lambda}\right)^{n/2} \frac 1{\sqrt{ \det (-H)}} \, e^{\lambda S(x_0)} \left[1+ O(1/\lambda)
\right].$$ where $H$ denotes the Hessian matrix of $S$. The real part of the action \eqref{action}  is given by
\be \mbox{Re} \, S = \sum_{a < b} \, j_{ab}^- \, \ln \, \frac{1}{2} ( 1 - {\mathbf  n}_{ab}^{-} \cdot {\mathbf
n}_{ba}^{-}) + j_{ab}^+ \, \ln \, \frac{1}{2} (1 - {\mathbf n}_{ab}^+ \cdot {\mathbf n}_{ba}^+), \ee where
${\mathbf n}_{ab}^\pm:=X_a^\pm{\mathbf  n}_{ab}$ and we have used the expression of the  inner product between
coherent states and all  phases have been absorbed in the imaginary part of the action. The maximality equation
and the critical equation (obtained by using standard $\SU(2)$ coherent state identities on the first variation
formula), become respectively:   \bea
\label{Re} X_a^{\pm } {\mathbf  n}_{ab} &=& - X_b^{\pm } {\mathbf  n}_{ba}\\
\label{critical} \sum_{b\colon b \neq a} k_{ab} \,\, {\mathbf  n}_{ab} &=& 0 \eea for all $a=1,...,5$. The second
one, implying closure of the coherent tetrahedra, is redundant, since we already chose our states to be of this
form.

\section{Bivectors, Gluing, and Boundary data}
\paragraph*{Bivectors and Gluing}
Now that we have the stationarity equations, the programme is to input them back into the action and give them a
geometric interpretation, \`a la Ponzano-Regge. The first obstacle is that these equations involve basically
3-dimensional rotations acting on vectors, but we would like to give them an
 interpretation of 4-dimensional geometry.

What one does  first is to regard the coherent tetrahedra as immersed in $\R^4$. Let us say lying in the plane
$x_0=0$, i.e. in the plane orthogonal to ${\bf e}_0=(1,0,0,0)$. Now, we immerse the vectors normal to the
associated triangles of the tetrahedron, $k_{ab}{\bf n}_{ab}\in\R^3$, by canonically associating them to bivectors
$B_0(k_{ab}{\bf n}_{ab})=(b_0^-({\bf n}_{ab}),b_0^+({\bf n}_{ab}))$ given in the `(self-dual, anti-self-dual)'
decomposition. Namely, we map $k_{ab}{\bf n_{ab}}\mapsto k_{ab}({\bf n_{ab}},{\bf n_{ab}})$, which is a simple
bivector and still lies in ${\bf e}_0^\bot$. By acting on this bivector with $G=(X_a^-,X_a^+)$ we get
\be\label{bivectors} B_{ab}:= k_{ab} \, (X_{a}^-,X_{a}^+)({\mathbf n}_{ab}, {\mathbf n}_{ab}). \ee which now lies
in the plane orthogonal to $G_a{\bf e}_0$,  and is still simple, since $|b^+|=|b^-|$. Since for all $b\neq a$, the
$B_{ab}$  lie in the same hyperplane, $(G_a{\bf e}_0)^\bot$, it can be shown that the set of bivectors satisfy the
so called `cross-simplicity constraints'  \cite{livine-2007-76} as well.

It can furthermore easily  be shown that  these bivectors satisfy all but one of the `bivector geometry
conditions'. By the {\em reconstruction theorem} in \cite{barrett-1998-39}, the full set are the conditions that
allows the set of bivectors to determine a unique non-degenerate geometric 4-simplex $\sigma$ in $\R^4$ (defined up to translation
and inversion).
The condition not yet satisfied is non-degeneracy, in the sense that for six triangles sharing a common vertex
we do not know if the six bivectors are linearly independent.

To address this, we must go back to the stationarity equations, \eqref{critical} and \eqref{Re}. Given a set
$\mathcal{B}=\{ \mathbf n_{ab} , k_{ab} \}_{a \neq b}$ of boundary data (with phases of coherent states still
undetermined) satisfying \eqref{critical}, suppose there exists two  sets  of five $\SU(2)$
 elements $\{U^+_a\}$ and $\{U_a^-\}$ which solve $U_a {\mathbf  n}_{ab}= - U_b {\mathbf  n}_{ba}$. Suppose
 furthermore that the solutions are distinct (not related by a global
symmetry) $\{U^+_a\}\sim\{U_a^-\}$.
 Then, equating $\{X_{a}^-,X_{a}^+\}=\{U^-_a,U_a^+\}$ it is straightforward to prove that indeed the
 bivectors defined in \eqref{bivectors}
are non-degenerate
  (see  Lemma 3 of \cite{Barrett:2009gg}). Hence, by
  the reconstruction theorem of bivectors,  they determine a unique geometric 4-simplex; the
  unique one (up to translation and inversion)
  \footnote{It can be shown that equating instead $\{X_{a}^\pm\}=\{U_a^\mp\}$ yields
  the 4-simplex with opposite orientation.} compatible with the boundary tetrahedra given by
  $\{ \mathbf n_{ab} , k_{ab} \}_{a \neq b}$.

   It then follows  that
  we can have at most two distinct sets of solutions $\{U_a\}$, since, were there a third set, we would, by the
  reconstruction  theorem, be able to generate a distinct geometric 4-simplex compatible with the
  same boundary tetrahedra.

\paragraph*{Boundary data}
Let us {\em focus on the case where the boundary data allows two solutions} $\{U^-_a,U_a^+\}$. Then, by the
reconstruction theorem we have that the geometry and orientation of  the triangles $\tau_{ab}$ and $\tau_{ba}$
(associated to the coherent tetrahedra $\tau_a$ and $\tau_b$ resp.) must be a priori compatible. Therefore we have
that there exists a {unique} $\hat g_{ab}\in\SO(3)$ for which  we have
 \bea\label{g_ab} \hat g_{ab} ( \tau_{ab} )&=& \tau_{ba} \nn\\
 \hat g_{ab} \mathbf n_{ab}&=& -\mathbf n_{ba}.\eea
This is called Regge-like boundary in \cite{Barrett:2009gg}.

Furthermore, in the discussion on coherent states we saw that we were left with an arbitrary $\U(1)$ rotation to
determine.
 The choice of phase for the boundary state above is given by picking the phase  $|k_{ab},\mathbf
n_{ab}\ra_{\mbox{\tiny R}}$ for $\tau_a$ to be arbitrary, and then fixing the phase of the state for the
corresponding triangle in $\tau_b$ to be
 \be\label{Regge}|k_{ab},\mathbf
n_{ba}\ra_{\mbox{\tiny R}}= g_{ab}J|k_{ab},\mathbf
n_{ab}\ra_{\mbox{\tiny R}}.\ee
we denote this choice by the sub-index R (or Regge). Regge-like
boundary data together with this choice of phase for the boundary state is called a {\em Regge state}.

\section{Dihedral angles and asymptotic formula}
The idea here is that the geometrically induced choice of
phases will correlate the value of the action at the critical points
with the dihedral angles.
\paragraph*{Dihedral angles}

Suppose $N_a$ is the outward unit normal vector to tetrahedron $a$. Then $N_a \wedge N_b$ defines a bivector which
is in the plane orthogonal to the triangle where tetrahedra $a$ and $b$ intersect. Therefore $*(N_a\wedge N_b)$
lies in the plane of the triangle; normalising it correctly then equals the definition of the bivector $B_{ab}$
\eqref{bivectors}. The dihedral rotation $\widehat D_{ab}\in\SO(4)$ is defined as the rotation that maps the
normal $N_a=G_a{\bf e}_0$ to the normal $N_b$ and stabilizes the orthogonal plane $N_a^\bot\cap N_b^\bot$ (the
plane of the bivector $B_{ab}$). These comments permit us to write\footnote{A bivector $N\wedge M=N\otimes M -
M\otimes N,$ as an element of the Lie algebra $\so(4)$, acts on vectors through the Euclidean metric inner
product. Also note that the isomorphism $\R^3\rightarrow\su(2)$ is effected through $v\mapsto v\cdot{\bf L}$.}:
 \bea \label{dihedral-simplex}
 D_{ab} &:=&  \exp \left(\Theta_{ab} \frac{N_b\wedge N_a}{|N_b\wedge N_a|} \right)\\ \smallskip
 \label{geometric bivectors} B_{ab} &=& k_{ab} * \frac{N_a\wedge
N_b}{|N_a\wedge N_b|}=k_{ab}(X_a^-,X_a^+)({\bf n}_{ab},{\bf n}_{ab})\eea Acting with the Hodge on
 \eqref{geometric bivectors} and using  $**=1$ and \eqref{dihedral-simplex} leads to \be D_{ab}=\left(\exp
\left(-\Theta_{ab}(\hat X_a^-{\bf n}_{ab})\cdot{\bf L}\right), \exp \left(\Theta_{ab}(\hat X_a^+{\bf
n}_{ab})\cdot{\bf L}\right)\right).\ee

 Now consider the following diagram:
\begin{equation}\label{commutative}
\xymatrix{\ar @{} [dr]  \tau_a \ar[d]_{(g_{ab},g_{ab})} \ar[rr]^{(X_a^-,X_a^+)} && ~\sigma_a \ar[d]^{D_{ab}}   \\
\tau_b\ar[rr]_{(X_b^-,X_b^+)} && ~\sigma_b  }
\end{equation}
where $\tau_a\subset\R^4$ are the tetrahedra at ${\bf e}_0^\bot$, and
 $\sigma_a\in\R^4$ are the actual geometrical ones in the 4-simplex $\sigma$ and the $g_{ab}\in \SU(2)$,
are defined in \eqref{g_ab}. Note that by the reconstruction theorem, the maps in the  diagram commutes when
acting on both the triangles $\tau_{ab}$ and on the internal normals $\mathbf n_{ab}$, hence (since  all maps in
the diagram are orientation preserving) the $\SO(4)$ action of the maps in the diagram commutes. Thus acting with
$\left((X_a^-)^{-1},(X_a^+)^{-1}\right)$ on to the left of the commuting diagram equation one gets the two
equations: \bea \label{glue1}(X_a^\pm)^{-1}X_b^\pm g_{ab}&=& \exp\left(\mp i\Theta_{ab}{\bf n}_{ab}\cdot
L\right)\eea

Now, as we know, the critical points satisfy closure and the conditions $ (X_a^{\pm})^{-1}X_b^{\pm} ( {\mathbf
n}_{ba}) = -{\mathbf n}_{ab}, $ for all $a\ne b$. The lift of this equation to the coherent states involves a
phase \be \label{critcoherent}(X_a^{\pm})^{-1}X_b^{\pm} | \mathbf n_{ba} \ra = e^{i \phi_{ab}^{\pm}} | -\mathbf
n_{ab} \ra. \ee Then, taking the inner product with $\la -\mathbf n_{ab}|$, with the Regge phase choice (paying
special attention to the indices and signs): \bea e^{i \phi_{ab}^{\pm}}&=& _{\mbox {\tiny R}}\la -\mathbf n_{ab}|
(X^\pm_a)^{-1}X^\pm_b |   \mathbf
n_{ba} \ra_{\mbox {\tiny R}} \\
\label{2}~&=& _{\mbox {\tiny R}}\la -\mathbf n_{ab}|(X_a^\pm)^{-1}X_b^\pm g_{ab}|  - \mathbf n_{ab}\ra _{\mbox
{\tiny R}}=
\overline{\la \mathbf n_{ab}|\exp\left(\mp\Theta_{ab}{\bf n}_{ab}\cdot L\right)|   \mathbf n_{ab} \ra_{\mbox {\tiny R}}}\\
~&=& e^{\pm \frac{i}{2}\Theta_{ab}} \eea where in the first equality of \eqref{2} we have used the Regge phase
choice \eqref{Regge}, and in the second equality of the same line we have used \eqref{glue1} and the properties of
the $J$ map.

Finally, by a simple direct computation, \eqref{action} becomes: \be\label{nondegenerateaction} S=\gamma\sum_{a<b}
k_{ab}\Theta_{ab}.\ee

 Even for a  boundary state $\{ |k_{ab}, \mathbf
n_{ab}\ra_{\mbox{\tiny R}}\}_{a \neq b}$ such that $\{U^+_a\}\nsim\{U_a^-\}$, we may still form degenerate
solutions of the form $\{X^\pm_a\}\sim\{U^+_a\}$ or $\{X^\pm_a\}\sim\{U^-_a\}$. These will contribute with the
 strength $\pm\sum_{a<b} k_{ab}\Theta_{ab}$ to the asymptotic formula.

Thus for non-degenerate boundary data we write for the total amplitude from  \eqref{formula1} and \eqref{action}:
\begin{multline} f_4(\{\lambda k_{ab},|   \mathbf
n_{ba} \ra_{\mbox {\tiny R}} \} ) \simeq \left(\frac{2 \pi}{\lambda}\right)^{12} \left[ 2 N^\gamma_{+-}\cos \left(
\lambda\gamma\sum_{a<b}k_{ab}\Theta_{ab}\right)
\right. \\
+
N^\gamma_{++}\exp{\left(i \lambda\sum_{a<b}k_{ab}\Theta_{ab}\right)}
+
\left.
N^\gamma_{--}\exp{ \left( -i\lambda\sum_{a<b}k_{ab}\Theta_{ab}\right)}\right]
\end{multline}
where the $N$'s are prefactors depending on the determinant of the Hessian but not on $\lambda$. {As in the case
of the Ponzano-Regge model, a cosine term appears because simplex geometries with either of the two possible
orientations can occur.}
\section{Conclusion}
We have studied the semi-classical limit of the four-simplex  amplitude of the Euclidean EPRL model for $\gamma<1$
and maximal set of solutions for the stationarity equations. The asymptotic formula contains the cosine of the
Regge action. However our asymptotic formula also contains two additional terms,  with exponentials of the same
Regge action formula, but without the Immirzi parameter $\gamma$. Interestingly, in the asymptotic limit, all of
these terms scale with the same exponent of the asymptotic parameter $\lambda$.

\end{document}